# Choroidal Vessel Segmentation on Indocyanine Green Angiography Images via Human-in-the-Loop Labeling

**Authors**

Ruoyu Chen, MD[1], Ziwei Zhao, MD[1], Mayinuer Yusufu, MBBS[4,5], Xianwen Shang, PhD[1], Danli Shi, MD, PhD[1-2*], Mingguang He, MD, PhD[1-3]

**Affiliations**

1. School of Optometry, The Hong Kong Polytechnic University, Kowloon, Hong Kong SAR, China

2. Research Centre for SHARP Vision, The Hong Kong Polytechnic University, Kowloon, Hong Kong SAR, China

3. Centre for Eye and Vision Research (CEVR), 17W Hong Kong Science Park, Hong Kong SAR, China

4. Centre for Eye Research Australia, Royal Victorian Eye and Ear Hospital, East Melbourne, Australia

5. Department of Surgery (Ophthalmology), The University of Melbourne, Melbourne, Australia

**Correspondence**

**\*Dr. Danli Shi,** MD, PhD., The Hong Kong Polytechnic University, Kowloon, Hong Kong SAR, China.

Email: danli.shi@polyu.edu.hk

# ABSTRACT

Human-in-the-loop (HITL) strategy has been recently introduced into the field of medical image processing. Indocyanine green angiography (ICGA) stands as a well-established examination for visualizing choroidal vasculature and detecting chorioretinal diseases. However, the intricate nature of choroidal vascular networks makes large-scale manual segmentation of ICGA images challenging. Thus, the study aims to develop a high-precision choroidal vessel segmentation model with limited labor using HITL framework. We utilized a multi-source ICGA dataset, including 55° view and ultra-widefield ICGA (UWF-ICGA) images for model development. The choroidal vessel network was pre-segmented by a pre-trained vessel segmentation model, and then manually modified by two ophthalmologists. Choroidal vascular diameter, density, complexity, tortuosity, and branching angle were automatically quantified based on the segmentation. We finally conducted four cycles of HITL. One hundred and fifty 55° view ICGA images were used for the first three cycles (50 images per cycle), and twenty UWF-ICGA images for the last cycle. The average time needed to manually correct a pre-segmented ICGA image per cycle reduced from 20 minutes to 1 minute. High segmentation accuracy has been achieved on both 55° view ICGA and UWF-ICGA images. Additionally, the multi-dimensional choroidal vascular parameters were significantly associated with various chorioretinal diseases. Our study not only demonstrated the feasibility of the HITL strategy in improving segmentation performance with reduced manual labeling, but also innovatively introduced several risk predictors for choroidal abnormalities.

# Introduction

The Choroidal vascular network is crucial in supplying oxygen and nourishment to the outer retina. Abnormalities of choroidal circulation may result in progressive dysfunction of retinal pigment epithelium (RPE) and photoreceptors, which have been increasingly recognized as critical factors in developing several chorioretinal diseases.[1] Previous studies demonstrated that pachychoroid spectrum diseases, such as central serous chorioretinopathy (CSC) and polypoidal choroidal vasculopathy (PCV), were associated with choroidal thickening and choroidal vascular morphological alterations.[2,3] Additionally, reduced perfusion in choroidal blood flow correlates with the thinning of the choroid and myopia severity.[4,5] Quantification of choroidal vasculature alterations may enhance our understanding of pathological processes and facilitate tailored management of chorioretinal abnormalities.

In the past few years, several choroid-related imaging biomarkers have been introduced via manual segmentation or deep learning-based approaches for screening and monitoring chorioretinal conditions. Indocyanine green angiography (ICGA) stands as the gold standard examination for visualizing choroidal vascular structure with the aid of a contrast agent, offering detailed information on choroidal vascular network.[6] Recent ultra-widefield ICGA (UWF-ICGA) technique provides a comprehensive overview of peripheral choroidal vasculature, such as vortex vein.[7] Choroidal vascular characteristics of several chorioretinopathy conditions have been explored in previous studies using ICGA images, such as vortex vein engorgement, fusiform choroidal vein, and choroidal vascular density.[2,8-10] However, most of those studies relied on labor-intensive manual segmentation and conventional binarization, lacking repeatability and effectiveness. Besides, few studies have investigated automatic approaches for accurate choroidal vessel segmentation based on ICGA/UWF-ICGA images, which represent the key prerequisite for investigating high-precision and reliable choroidal biomarkers.

In the past few years, a large volume of studies have been conducted for achieving and optimizing automated vessel segmentation from fundus images using deep learning (DL).[11,12]

However, the majority of these systems primarily focus on the retinal slab rather than the more complex choroidal vascular layer. Several factors, such as the intricate nature of choroidal vascular networks and their fluorescence alterations, make large-scale manual segmentation of ICGA images challenging. Human-in-the-loop (HITL) strategies, incorporating low-shot human annotation and active learning as cornerstones, have been recently introduced into the field of image processing.[13,14] Integrating HITL into DL model training could reduce the time-consuming manual segmentation process, allowing humans to make necessary corrections to preliminary segmentation results. Collaboration between machine intelligence and human feedback could benefit training efficiency and model performance.[15,16]

This study aims to develop an automated algorithm to segment choroidal vessels on ICGA/UWF-ICGA images and calculate vascular parameters by employing a DL framework and HITL labeling. This model is expected to achieve high-precision measurements of choroidal vascular morphologies and provide reliable biomarkers for the better management of chorioretinal conditions.

## Methods

### Datasets

We utilized a multi-source ICGA image dataset, consisting of 55° view and ultra-widefield ICGA (UWF-ICGA) images, to develop this choroidal vessel segmentation model. These ICGA images were captured from patients with chorioretinal diseases, including CSC, PCV, pathological myopia (PM), age-related macular degeneration (AMD), choroidal neovascularization (CNV), and ocular inflammatory diseases, etc., as well as those with no apparent abnormalities on ICGA images (classified as normal participants in this study). All patient data underwent anonymization and de-identification processes. To develop a precise segmentation model for the choroidal vascular network, we selected ICGA images captured between 30 seconds and 3 minutes after dye injection, as these early-phase images provide the most explicit visualization of the choroidal vasculature. The ICGA images were acquired using Heidelberg Spectralis cameras (Heidelberg, Germany) at a resolution of 768×768.

**Segmentation Model Development**

The interactive HITL was employed for developing choroidal vessel segmentation model. Firstly, we used a pre-trained vessel segmentation model[17] for choroidal vessel pre-segmentation on ICGA images. The model could segment retinal vessel across multimodality images. The 55° view ICGA and UWF-ICGA images were resized to 512×512 and 1024×1024 respectively, then fed into this pre-trained model. In addition, we adjusted the threshold to achieve high-sensitivity vessel prediction, enabling the segmentation of visible vascular structure on ICGA images. Secondly, using a custom software (VesselLabel)[12], two experienced ophthalmologists (R.C and Z.Z) randomly selected the pre-segmented vessel maps for manual modification. The pre-segmented choroidal vessel map could be overlaid on the original ICGA images or the contrast-enhanced ones. The ophthalmologists could switch between different modes and correct vessel annotations proposed by the pre-segmentation model, such as removing false labels (i.e., retinal vessel) and supplementing vessels that the pre-segmentation model overlooked. Thirdly, we trained a generative adversarial network (GAN) model[18] to segment choroidal vessel from ICGA by using these human-modified choroidal vessel maps as label for 50 epochs. Subsequently, we applied this refined model to new ICGA images to get predictions as the 2$^{nd}$ version of choroidal pre-segmentation, followed by further corrections. The iterative steps were repeated until the quality of predicted choroidal vessel maps reached satisfaction under visual inspection. The flowchart of this study is shown in **Figure 1.** The models were trained with a batch size of 4 and a learning rate of 0.0002.

**Accuracy of Vessel Segmentation**

Two experienced ophthalmologists (R.C and Z.Z) conducted visual quality evaluation. Dice coefficient was used to assess the consistency between annotators. We assessed the accuracy of choroidal segmentation at the pixel level. Quantitative evaluation was conducted using F1-score, the area under the receiver operating characteristic curve (AUC), accuracy, sensitivity, and specificity, using human-rectified vessels (R.C.) as ground truth.

**Quantitative vessel metrics and their associations with chorioretinal diseases**

The Retina-based Microvascular Health Assessment System (RMHAS) pipeline was employed to quantify vessel characteristics from the predicted choroidal vessel maps across five dimensions: density, complexity, tortuosity, caliber, and branching angle.[12] Since the vascular trees were segmented into numerous vessel segments for analysis, resulting in hundreds of individual measurements for each metric per image, we utilized summary statistics—mean, standard deviation, maximum, and minimum values to consolidate these measurements and provide a comprehensive overview of each image. These summary statistics offer a multifaceted representation of the measurements.[19-21] Additionally, to evaluate the reliability and clinical significance of these automatically calculated parameters on segmented choroidal vascular maps, we compared these parameters in normal participants (without any abnormalities on ICGA images) and those with pachychoroid spectrum diseases (CSC, PCV) and choroidal thinning disease (PM).

**Statistical Analysis**

Continuous variables are reported as mean ± standard deviation (SD), while categorical variables are presented as frequency and percentage. We removed extreme outliers for choroidal vascular parameters using the Robustbase package in R (range=3), accounting for skewness. To enable a more uniform comparison across variables and eliminate scale differences, we normalized the values of choroidal vascular parameters to SD units. Multivariable logistic regression models were employed to investigate the association between choroidal vascular parameters and CSC, PCV, and PM, with adjustments for age and sex. Given the multiple parameters and consequent multiple comparisons, P values were adjusted for the false discovery rate (FDR) to control for type I error for multiple comparisons, and the significance level was defined as a two-tailed P value of 0.05. All statistical analyses were conducted using R version 4.3.3.

## Results

**Data Demographics**

We finally repeated four cycles of above iterative process. One hundred and fifty 55° view

ICGA images were used for the first three cycles (50 images per cycle), and twenty UWF-ICGA images for the last cycle. The average time needed to manually correct a pre-segmented ICGA image in each cycle was 20 minutes, 5 minutes, and 1minutes for 55° view ICGA images and 10 minutes for UWF-ICGA images, respectively. Additionally, we retrospectively included ICGA images from 394 patients with complete ICGA reports and definitive diagnoses, comprising 151 normal participants, 46 CSC participants, 130 PCV participants and 67 PM participants. The median age of the participants was 50.03 (±19.43) years, and 235 (59.64%) were male. Detailed characteristics of diseases spectrum are summarized in **Table 1**.

**Accuracy of Choroidal Vessel Segmentation**

The ground truth was established using manually segmented choroidal vessel maps by two ophthalmologists (R.C. and Z.Z.). The intra-grader Dice coefficient was 0.870. The predicted choroidal vascular maps showed good alignment with the ground truth, using R.C.'s label as reference: AUC 0.975 (95%CI: 0.967-0.983), F1-score 0.867 (95%CI: 0.856-0.878), accuracy 0.950 (95%CI: 0.946-0.954), sensitivity 0.858 (95%CI: 0.844-0.872), specificity 0.972 (95%CI: 0.968-0.976). The model also demonstrated high accuracy on segmenting choroidal vascular on UWF-ICGA images: AUC 0.937 (95%CI: 0.914-0.960), F1-score 0.780 (95%CI: 0.763-0.797), accuracy 0.895 (95%CI: 0.862-0.928), sensitivity 0.784 (95%CI: 0.762-0.806), specificity 0.927 (95%CI: 0.895-0.959). The examples of predicted choroidal vessel networks on ICGA and UWF-ICGA images are shown in **Figure 2** and **Figure 3**.

**Choroidal vascular measurements and chorioretinal diseases**

RMHAS extracted 164 choroidal vessel measurements on the predicted choroidal vessel segmentation, covering five categories: density, complexity, tortuosity, caliber and branching angle. After removing measurements with 90% missing data or over 95% of elements were the same value, 102 choroidal vessel measurements were included in further analysis. The significant associations (FDR-adjusted p value < 0.05) between various chorioretinal diseases and multiple vessel measurements are illustrated in **Figure 4**. All the results were adjusted for sex and age.

The relationship of various choroidal vessel measurements with CSC are demonstrated in **Figure 4A**. Twenty-six parameters significantly associated with CSC after adjusting for sex and age. The complexity-related parameters (Strahler number), most of the caliber-related parameters (mean caliber, minimum caliber, maximum caliber, caliber range, and terminal caliber), and tortuosity-related parameters (angle-based tortuosity, curve angle) were significantly and positively associated with CSC. On the contrary, most of the choroidal vascular density-related parameters (arc length, chord length, and vessel skeleton density), tortuosity-related parameters (tortuosity density, fractal tortuosity, and inflection tortuosity), and all the branching angle-related parameters (angular asymmetry) were negatively correlated with CSC. Specifically, branching density (mean value per image) was associated with increased risk of CSC (odds ratio [OR] = 1.72 [95% CI: 1.12-2.63]), angle-based tortuosity (mean value per image) was also related to increased risk of CSC (odds ratio [OR] = 1.77 [95% CI: 1.16-2.68]. Additionally, mean width of choroidal vascular (mean value per image) is associated with increased risk of CSC (odds ratio [OR] = 2.24 [95% CI: 1.53-3.84]), max width of choroidal vascular (mean value per image) is associated with increased risk of CSC (odds ratio [OR] = 2.26 [95% CI: 1.41-3.61]).

The association of multiple choroidal vessel parameters and PCV are illustrated in **Figure 4B.** Twenty-eight parameters showed a significant association with PCV after adjusting for sex and age. Among these results, most complexity-related metrics (Strahler number, level, and number of terminal points), caliber-related metrics (mean caliber, maximum caliber, minimum caliber, and caliber range), branching angle-related metrics (branching angle) and all tortuosity-related metrics were associated with increased risk of PCV. In contrast, most choroidal vascular density-related parameters (arc length, chord length, and vessel skeleton density) were inversely correlated with PCV. To be more specific, branching density (mean value per image) is associated with increased risk of PCV (odds ratio [OR] = 1.91 [95% CI: 1.34-2.74]), complexity-related metrics, Strahler number (mean value per image) is associated with increased risk of PCV (odds ratio [OR] = 2.10 [95% CI: 1.43-3.00]), angle-based tortuosity (mean value per image) is associated with increased risk of PCV (odds ratio [OR] = 1.86 [95% CI: 1.32-2.61]. Besides, mean width of choroidal vascular (mean value per image) is associated with increased risk of PCV (odds ratio [OR] = 1.65 [95% CI: 1.15-2.36]), max width of choroidal vascular (mean value per image) is associated with increased risk of PCV (odds ratio [OR] = 1.57 [95% CI: 1.09-2.27]).

The significant relationships between multiple choroidal vessel parameters and PM are illustrated in **Figure 4C.** Twenty-nine parameters showed a significant association with PM after adjusting for age and sex. Among these results, density-related parameters (arc length and chord length), complexity-related parameters (Strahler number, level, and fractal dimension), caliber-related parameters (mean caliber, maximum caliber, minimum caliber, surface area, caliber range and length diameter ratio) and branching angle-related parameters (Asymmetry Ratio) were positively correlated with the occurrence of PM. On the contrary, most of the tortuosity-related parameters (tortuosity, tortuosity density, curve angle, curve angle tortuosity and inflection tortuosity) were negatively correlated with PM. Arc length (mean value per image) is associated with increased risk of PM (odds ratio [OR] = 6.46 [95% CI: 3.28-12.70]), Strahler number (mean value per image) was associated with increased risk of PM (odds ratio [OR] = 2.73 [95% CI: 1.84-4.08]), fractal dimension (mean value per image) was associated with increased risk of PM (odds ratio [OR] = 3.28 [95% CI: 2.15-5.01]), asymmetry ratio (mean value per image) is associated with increased risk of PM (odds ratio [OR] = 1.51 [95% CI: 1.11-2.06]). Besides, the mean width of choroidal vascular (max value per image) is also positively correlated to PM (odds ratio [OR] = 1.70 [95% CI: 1.21-2.39])

## Discussion

In the present study, we innovatively developed a high-precision choroidal vessel segmentation model with limited labor by leveraging the power of HITL framework. This model achieved high accuracy both on 55° view ICGA and UWF-ICGA images. These auto-quantified multidimensional parameters on segmented choroidal vessel maps were associated with chorioretinal conditions, serving as potential biomarkers for risk prediction of CSC, PCV, and PM. Our study not only demonstrated the feasibility of the HITL strategy in improving segmentation performance with reduced manual labeling, but also introduced several risk predictors for choroidal abnormalities.

A well-developed image processing model requires massive training data with accurate annotation. However, labeling large-scale samples is labor-intensive and time-consuming. Since humans have rich prior knowledge, encouraging algorithms to interact with experienced specialists helps to improve performance in various tasks.[22] Inspired by this theory, the HITL framework has recently received significant attention. This novel framework introduces an interactive object detection architecture involving humans in correcting a few annotations suggested by a pre-trained detector.[23,24] This method avoids a cold start by incorporating essential knowledge and experience into the pre-trained learning system via limited iterative labeling. In the present study, the HITL architecture substantially decreased the time required for manual labeling cycle by cycle. It contributed to achieving high accuracy in predicting choroidal vascular maps using a relatively small number of training samples.

More recently, many studies have investigated the relationship between choroidal vascular alterations and various chorioretinal diseases. Optic coherence tomography (OCT) B-scan and enface projection images are common image modalities for capturing choroidal vascular-related biomarkers within macular regions, such as choroidal vascularity index (CVI) on B-scan images, flow voids, and choroidal capillary density on enface images.[25-27] However, limited scanning area and potential artifacts may affect the comprehensiveness and accuracy of quantification. ICGA holds an essential role in imaging choroidal circulation and is capable of providing more information about vascular morphologies, especially in early-phase images. The extraction of choroidal vascular features mainly relied on time-consuming manual segmentation.[10,28] To address this issue and achieve more accurate evaluation, we presented this automatic choroidal vascular segmentation model that enables effective quantification of vascular characteristics on 55° ICGA and UWF-ICGA images, including density, complexity, tortuosity, branching angle, and caliber. This pioneering algorithm represents a promising approach for the multidimensional evaluation of complex choroidal vascular networks.

Choroidal venous overload is a major characteristic in CSC pathogenesis, with important treatment and prognostic implications.[29] Enlarged choroidal vessels and choroidal intervortex venous have been frequently observed on ICGA images of CSC patients.[30,31] Similarly, PCV is another phenotype of pachychoroid diseases spectrum, characterized by dilated choroidal veins and choroidal vessels remolding.[32,33] The proposed model successfully generated several quantitative parameters based on the predicted choroidal vascular segmentation. The results showed that most parameters related to caliber, complexity, and tortuosity were significantly correlated with an increased risk of CSC and PCV, consistent with previously reported pathological processes. Moreover, in the current study, PM eyes demonstrated larger choroidal vascular diameters, possibly due to increased visibility of choroidal large vasculature resulting from axial elongation-associated choriocapillaris loss.[34,35] These findings indicate that these quantitative metrics might be reliable imaging biomarkers with promising generalizability, facilitating an updated understanding of the pathological mechanisms behind pachychoroid diseases in the future.

## Limitations

Nevertheless, this study also has some limitations. Firstly, the study utilized an in-house ICGA image dataset, with all participants coming from the Chinese region. Therefore, large-scale ICGA databases covering multi-ethnic populations are needed to evaluate the robustness of our algorithm. Secondly, although these vascular measurements have been validated across three common chorioretinal diseases, further validation is necessary to encompass a wide range of diseases. Thirdly, ICGA is an invasive procedure, and future investigations on non-invasive ICGA-like choroidal vessel quantification are promising using generative artificial intelligence.[36,37]

## Conclusions

In summary, the current study represents the first attempt to develop an automatic choroidal vascular segmentation model via HITL strategies. Moreover, the proposed algorithm can quantify choroidal vascular features from multiple perspectives, with reliable performance validated across three common chorioretinal diseases. These findings underscore the potential of applying this choroidal vascular segmentation model for accurate and comprehensive analysis of choroidal vascular alteration in multiple diseases.

## Data and code availability

The data used for model development of this study are not openly available due to reasons of privacy. The authors do not have the permission to distribute the dataset publicly. Code is available at: https://github.com/NVIDIA/pix2pixHD

## Ethics

We utilized de-identified existing data for our study, which received approval from the Institutional Review Board of the Hong Kong Polytechnic University.

## Funding

D.S. and M.H. disclose support for the research and publication of this work from the Start-up Fund for RAPs under the Strategic Hiring Scheme (Grant Number: P0048623) and the Global STEM Professorship Scheme (Grant Number: P0046113) from HKSAR. The funders had no role in study design, data collection and analysis, decision to publish, or manuscript preparation.

## Conflicts of interest

M.H. and D.S. are inventors of the technology mentioned in the study patented as"A retinal chorioretinal vessel segmentation method based on the conditional generative adversarial network (CN114782339A)".

## Author contributions

D.S. conceived the study. D.S. built the deep learning model. D.S. and R.C conducted the literature search, analyzed the data. R.C. and Z.Z. performed manual correction on pre-segmented choroidal vascular maps and completed visual evaluation. R.C. wrote the manuscript. M.H., provided the data and facilities. All authors critically revised the manuscript.

**Figure 1 | Flowchart of this study.** ICGA=Indocyanine green angiography

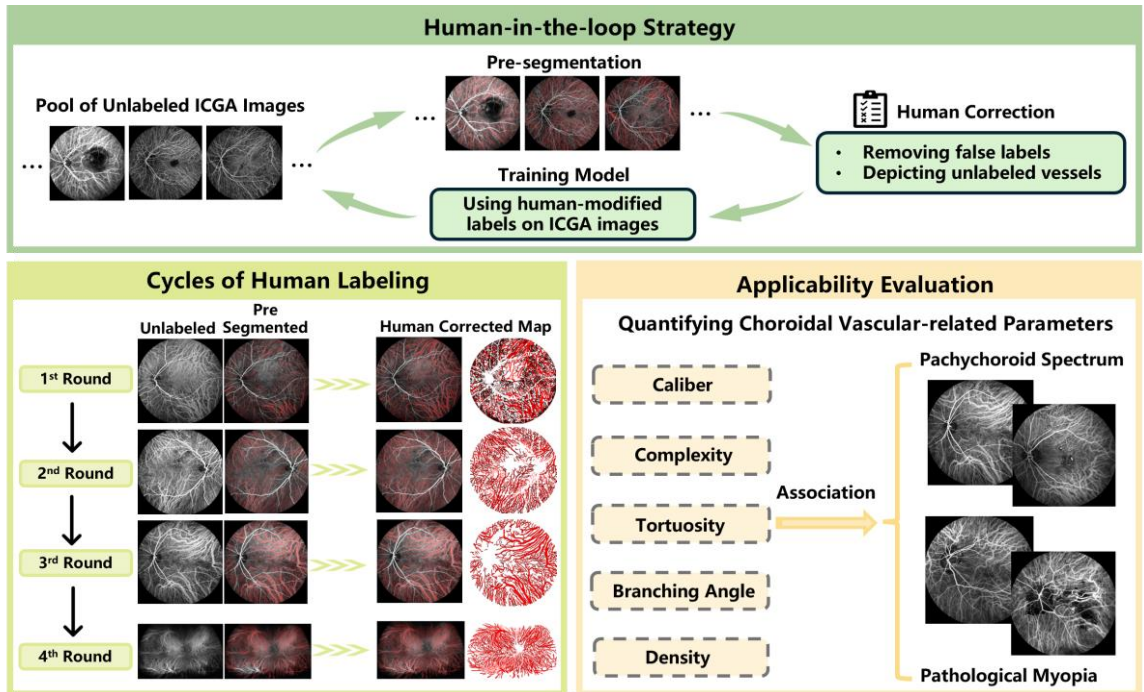

**Figure 2 | Examples of predicted choroidal vascular networks on indocyanine green angiography (ICGA) images and vascular measurement plots**

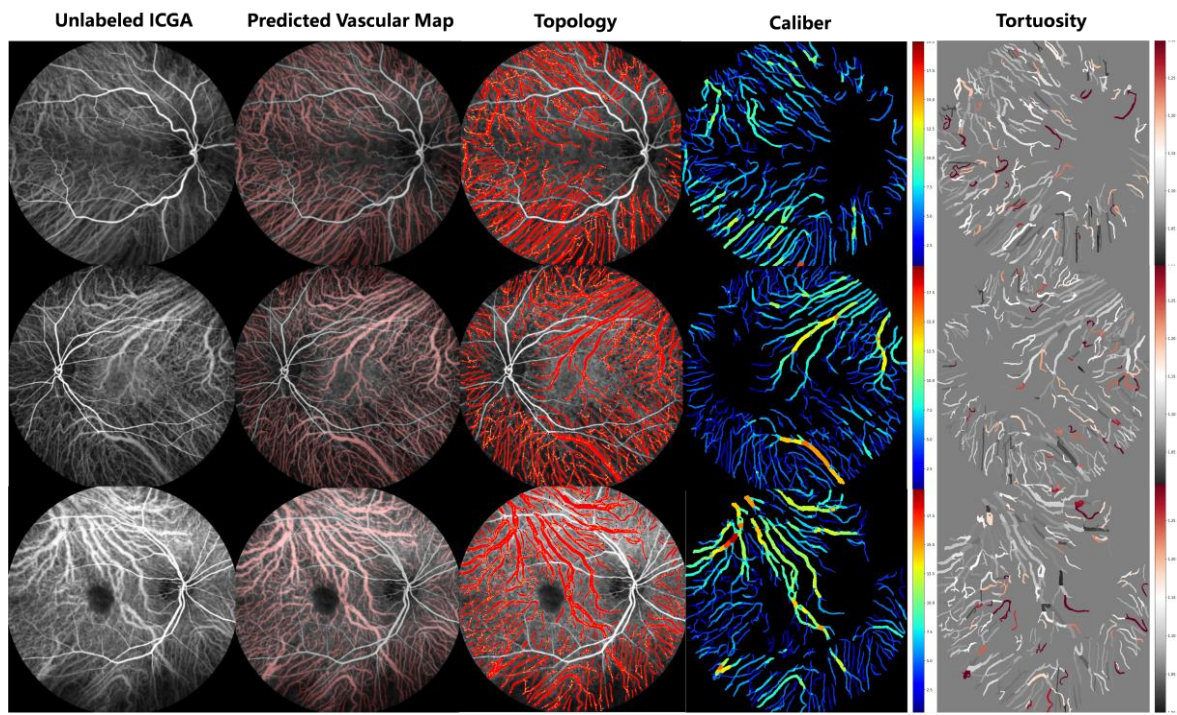

**Figure 3 | Examples of predicted choroidal vascular networks on ultra-widefield indocyanine green angiography (UWF-ICGA) images and vascular measurement plots**

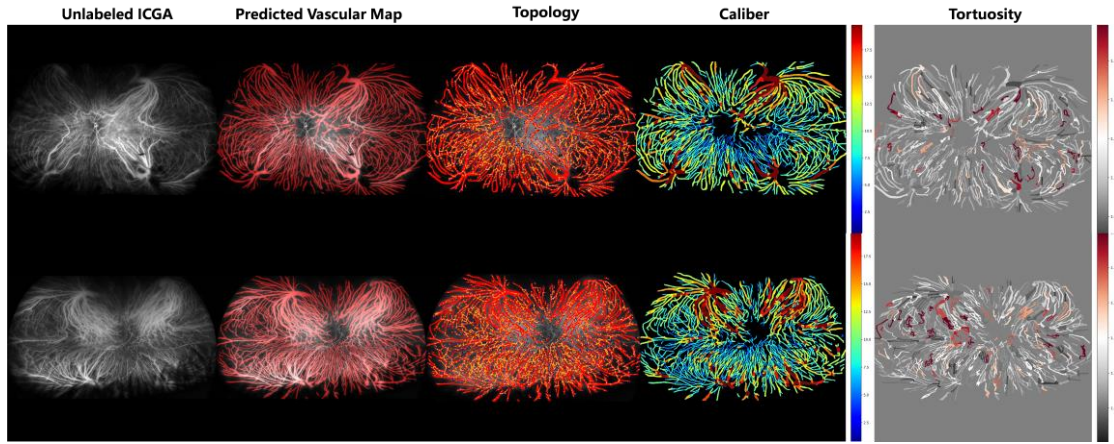

**Figure 4A | Association between central serous chorioretinopathy and selected choroidal vascular-related parameters. P values were adjusted for sex and age.**

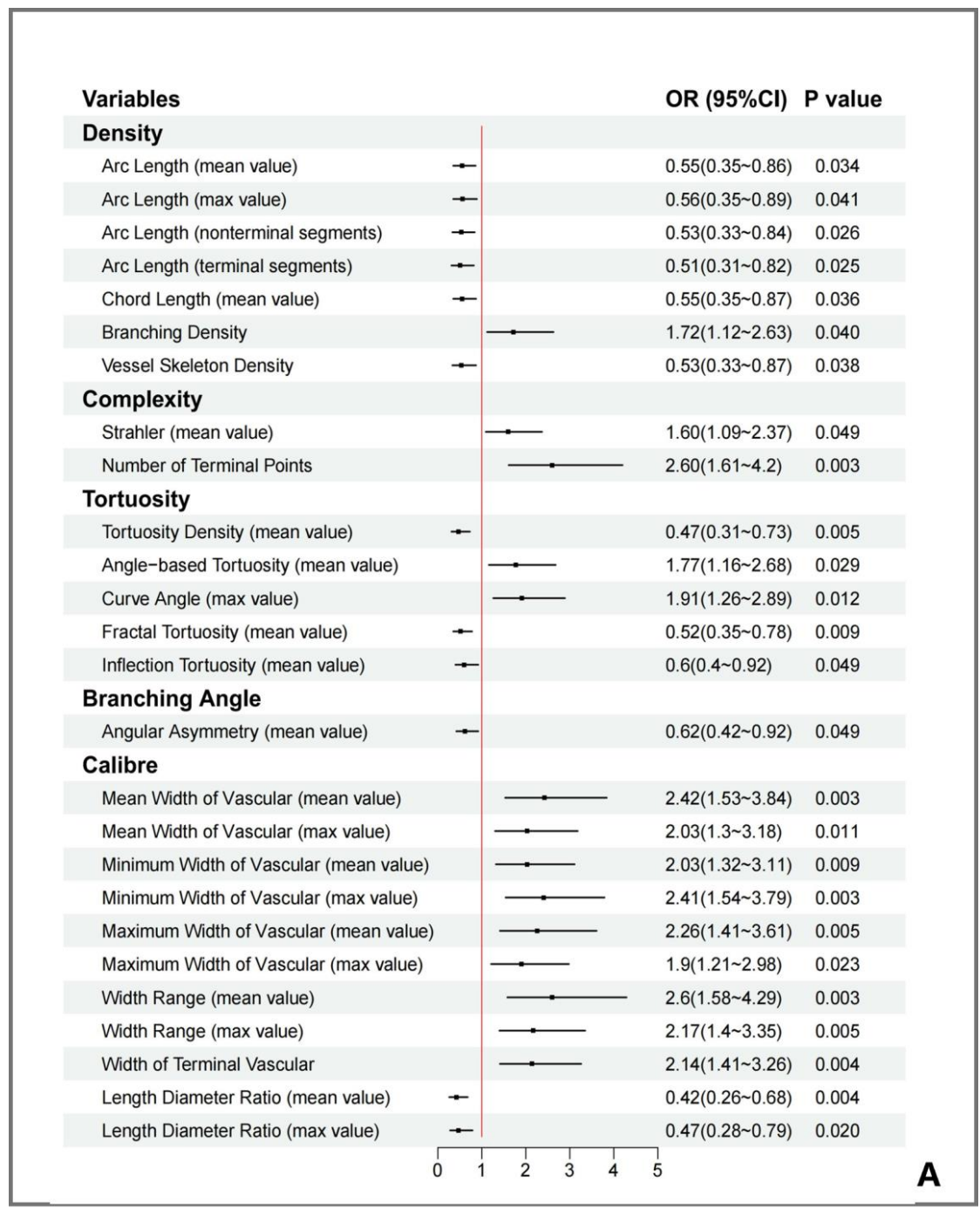

**Figure 4B | Association between polypoidal choroidal vasculopathy and selected choroidal vascular-related parameters. Pathological myopia. P values were adjusted for sex and age.**

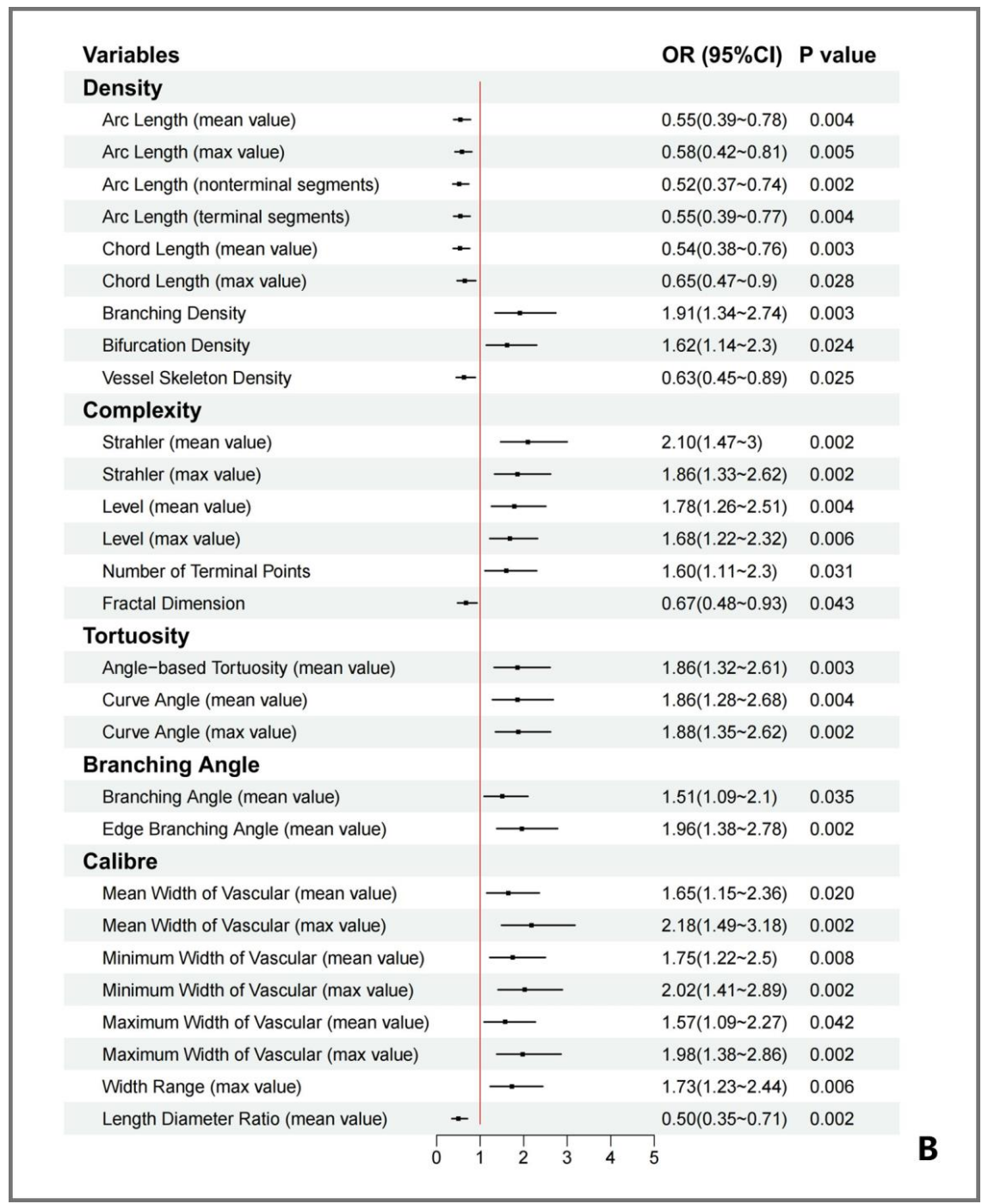

**Figure 4C | Association between pathological myopia and selected choroidal vascular-related parameters. P values were adjusted for sex and age.**

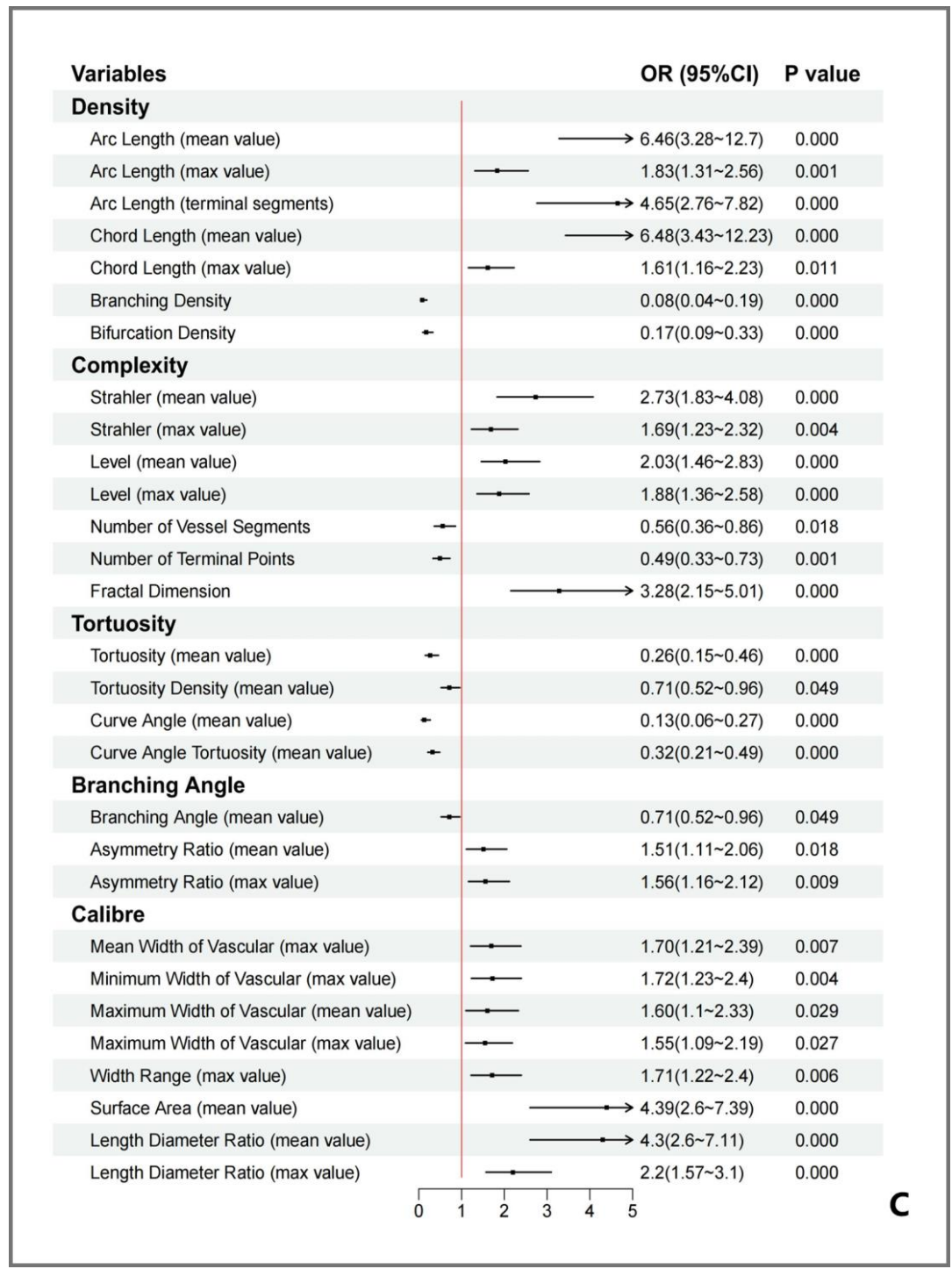

**Table 1 | Data Characteristics**

| Tasks | | | | | | | |
|---|---|---|---|---|---|---|---|
| **Model Development [Images, N (%)]** | | | | | | | |
| | Total | Normal | CSC | PCV | PM | AMD | Others |
| Round 1 | 50 (100%) | 6 (12%) | 9 (18%) | 11 (22%) | 7 (14%) | 6 (12%) | 11 (20%) |
| Round 2 | 50 (100%) | 6 (12%) | 7 (14%) | 10 (20%) | 9 (18%) | 8 (16%) | 10 (20%) |
| Round 3 | 50 (100%) | 5 (10%) | 10 (20%) | 7 (14%) | 15 (30%) | 5 (10%) | 8 (16%) |
| Round 4 | 20 (100%) | 3 (15%) | 1 (5%) | 1 (5%) | 0 (0%) | 4 (20%) | 11 (55%) |
| **Association Analysis [N (%)]** | | | | | | | |
| Patients | Total | Normal | CSC | PCV | PM | | |
| | 394 (100%) | 151 (38%) | 46 (12%) | 130 (33%) | 67 (17%) | | |